\documentclass[10pt]{article}
\usepackage[dvips]{graphicx}
\usepackage{graphicx}
\usepackage{epsfig}

\begin{document}
\title{\bf Crossover component in non critical dissipative sandpile models}
\author{A. Benyoussef, M. Khfifi and M. Loulidi$^1$\\
\it Laboratoire de Magn\'etisme et de Physique des Hautes Energies,\\
\it Departement de physique, Facult\'e des Sciences, B.P.1014, Rabat, Morocco\\
}       
\date{}
\maketitle
\begin{flushleft}                
\bf\large{Abstract}
\end{flushleft}
The effect of bulk dissipation on non critical sandpile models is studied using both multifractal and finite size scaling analyses. We show
numerically that the local limited (LL) model exhibits a crossover from multifractal to self-similar behavior as the control parameters
$h_{ext}$ and $\epsilon$ turn towards their critical values, i.e. $h_{ext} \rightarrow 0^+ $ and $\epsilon \rightarrow \epsilon_c$. The critical
exponents are not universal and exhibit a continuous variation with $\epsilon$. On the other hand, the finite size effects for the local unlimited (LU), non local limited (NLL), and non local unlimited (NLU) models are well described by the multifractal analysis for all values of dissipation rate $\epsilon$. The space-time avalanche structure is studied in order to give a deeper understanding of the finite size effects and the origin of the crossover behavior. This result is confirmed by the calculation of the susceptibility.

\vspace*{0.5cm}

\vspace*{6cm}
Published in Eur. Phys. J. B{\bf 43} (2005) 213-220
\renewcommand{\thefootnote}{}
\footnote{} 
\renewcommand{\thefootnote}{\arabic{footnote}}
\footnotetext[1]{Associate member of ICTP} 
\newpage
\section{Introduction}
\hspace*{0.5cm}Sandpiles are prototype out of equilibrium dynamical systems that usually present a self organized criticality(SOC).
This fundamental concept in modern physics of non-equilibrium phenomena was introduced by Bak, Tang, and Wiesenfeld [1-5] in order to
 explain the emergence of scaling behavior and fractal structure observed in nature. They proposed the SOC concept as one procedure to
 describe the basic mechanism that creates generic scale-free behavior [1,2]. This behavior emerges when an externally driven dissipative 
system organizes itself into a state where all spatial and temporal events are correlated over many orders of magnitude. The main feature of
the SOC is that its details are not determined by fine-tuning or initial conditions. Moreover, open boundary conditions or bulk dissipation
 insure a balance between input and output flow and allow to non-equilibrium stationary state [6-8]. The SOC is postulated be applicable over a wide variety of natural phenomena, spanning from microscopic to the astrophysical scale. Bak and his co-workers used the sandpile [1-4] model as a paradigm because of the crude analogy between its dynamical rules and the way sand topples when building a real sand pile. The sandpile model illustrates the SOC for a large class of complex systems [5 and references therein]: biology, economics, forest fire models, earthquakes, the game of life, invasion percolation together with systems that might be expected to exhibit self-similar and scale invariance behavior.\\
\hspace*{0.5cm}Sandpile systems are modeled as a regular array of columns consisting of cubic sand grains; the usual formalization 
considers each lattice site to be characterized by a state variable $h(i)$, where $h$ is the height or local slope of the sand column at a
giving site (i). Two mechanisms are crucial in such models: slow addition of new grains, which is simply performed by selecting either a random or a 
fixed column and increasing its height by one unit, and the relaxation process. In relaxation process, if the slope in column i exceeds a 
threshold value, then an amount of column sand is redistributed among its neighbors following a series of topples which may give rise to an 
avalanche that subside after a finite period of time. The avalanche size 'S' is giving by the number of toppling sites. In SOC, the response 
to an external perturbation results in avalanches of all sizes with power-law distributions of the form, $D(s, L)\sim s^{-\tau} g(s/L^{\it D})$,
where {\it L} is the system size. The critical exponents $\tau$ and {\it D}, depend on the model one defines. For some stochastic sandpile 
models, the critical exponent $\tau$ varies continuously [9] by varying the disorder in the system. It is believed that the two time-scale
 separation [6,10,11] (deposition and relaxation) and metastability are essential for the existence of scale invariance in these models. The
 dynamics of the sandpiles have been intensively studied both theoretically [6,12] and experimentally [13, 14], and some exact results were 
derived for abelian sandpile models[15]. Originally, experimental studies showed that the sandpile model leads to a clear disagreement with
 numerical simulations of theoretical models. Indeed, if the pile is tilted several degrees above the angle of repose grains start to flow and
 the system exhibits only a first order transition. On the other hand, rice-pile experiments in quasi one-dimensional systems display 
SOC with a power-law distribution of the avalanche size with critical exponent $\tau \approx 2.02$ [14], and this will occur depending on 
the detail of the grain level-dissipation.\\
\hspace*{0.5cm} Simple rules of the dynamics in sandpile models lead to simple equations according to mean-field theory, which is based on the
single site approximation of the master equation [6]. One may associate rates $ h_{ext}$ and $\epsilon$, respectively, with the addition and 
dissipation processes. The inverse of the parameter $h_{ext}$ is simply the typical waiting time between different avalanches, i.e.
$\tau_d \approx 1/h_{ext}$. In this case, the criticality, which results from non local interactions, is obtained by fine-tuning the control parameters
as in continuous phases transitions.\\
\hspace*{0.5cm}For a deeper understanding of the significance of SOC a connection to conventional critical points has been illustrated through some 
simple models[16]. As a result, it was shown that the SOC can be understood as an aspect of multiple absorbing state models since the
criticality is reached in the limit $h_{ext} \rightarrow 0$ and $\epsilon \rightarrow 0^+$ with  $ h_{ext}/\epsilon \rightarrow 0$ similar to Contact
 Process-like models, and the power-law avalanche distribution is found to be a general feature of  models with many absorbing configurations [17]. On the other hand, using an unified mean field theory the main prediction is that criticality is ensured by the divergence of the zero-field 
susceptibility which is giving by $\chi = \partial \rho_a/\partial h_{ext}$, where $\rho$ is the density of active sites. In the limit of vanishing
 control parameters, the stationary state displays scaling that is characteristic of SOC.\\
\hspace*{0.5cm}The effect of bulk dissipation has been studied in a two dimensional dissipative height sandpile model, and it was shown that the 
SOC behavior doesn't occur for discrete driven dissipative model while it was observed for continuous ones given a particular choice of the
dissipation rate~$\epsilon$[8].\\
\hspace*{0.5cm}Our aim is to study numerically the effect of the bulk dissipation on non critical one dimensional sandpile models that obey multifractal analysis, and show that in contrast to the previous dissipative systems, the dissipation may influence the avalanche dynamics leading to SOC for LL models. The outline of this paper is as follow: in sec. II we define the models, the methods used and explain
the problem of the multifractality, while sec. III is devoted to the study of the dissipation effects on non critical models defined in sec. II. In sec IV we give a general conclusion.
\section{Models and methods}
\hspace*{0.5cm}For our systems we assume integer heights $h(i)$ at lattice sites i=1,2...{\it L}, where  {\it L} is the size of the system.
 The local slope of the pile $z (i) $ at site ($i$) is defined as the height difference between two nearest neighbors:
\begin{equation}
z (i) =h(i)-h(i+1).
\end{equation}
The boundary conditions of the system are reorganized such that the grains can flow out of the system from the right side only. The system consists of a plate of length $ L$, with a wall at $i =0$ and an open boundary at $i =L+1$:
\begin{center}
                       $z (i) =0$,      if    i = 0\\
               	       $h (i) = 0$,     if    $ i > L$.
\end{center}
\rm The profile of the system evolves through deposition and relaxation. In deposition, a particle is added to a random site $i$ and
\begin{equation}
 h (i) \rightarrow  h(i)  +  1. 
\end{equation}
\rm During relaxation, we look at all unstable columns of the pile: a column i of the pile is considered active if $z (i) > z_c$. The number of grains N to be toppled is determined by either the limited or the unlimited rule. In the case of limited model, the number N is fixed while for the unlimited case, the number of toppled grains depends of the slope at the active sites therefore N = {\it z}(i)-M, where M is a fixed number. Some degrees of non conservation can be introduced in the model by allowing a dissipation of energy during relaxation events. In a discrete energy model, one can introduce a probability $\epsilon$ that the N transferred particles in the relaxing process are annihilated. The relaxation can take place either according to the local (L) or non local (NL) rule

\begin{eqnarray}
\nonumber
&&h(i+1) = h(i+1) + N ; \hspace*{3cm} (L)\\	\nonumber
&&h(i+j) =h(i+j) + 1  ,  \hspace*{0,5cm} j=1...N ; \hspace*{0,7cm} (NL) 
\end{eqnarray}

Thus four different 1D sandpile models [18] may be defined as follow: \\
1. Local limited model (LL),\\
2. Local unlimited model (LU),\\
3. Non local limited model (NLL),\\
4. Non local unlimited model (NLU).\\
\hspace*{0.5cm}In order to check the effect of dissipation under the total
size of avalanches $F$ (which is defined as the total number of toppling sites) and the number of grains that drop off the edge, $D$, we are interested in applying two techniques to analyze our data. In the first technique, where the system dynamics exhibit self-similarity behavior, the probability distribution function (PDF) $\rho(X, L)$ of the events $F$ and $D$ is fit to the finite-size scaling form represented by,
\begin{equation}
 \rho(X, L) =L^{-\beta}g(X/L^{\nu}),  
\end{equation}
where $\beta$ and $\nu$ are critical exponents associated with the distribution $\rho$ and the $ X$ quantity, respectively,  and $g$ is a universal scaling function. This form results from the fractal structure of the system that presents self-similar properties. Thus, the fractal analysis is successfully applied to describe the finite size effects of such systems which exhibit scale invariance.\\
In the other technique, the PDFs are better fit with the multifractal form [18-21] given by the $f-\alpha$ representation:
\begin{equation}
{\rm log_{10}}\rho(X, L)/{\rm log_{10}}(L/L_0)= f({\rm log_{10}}(X/X_0)/{\rm log_{10}}(L/L_0)).
\end{equation}
where $ X_0 $ and $ L_0 $ are constants that depend on the value of $\epsilon$ and $\alpha={\rm log_{10}}(X/X_0)/{\rm log_{10}}(L/L_0)$.
 In the case where $f$ is a linear function, the probability distribution is then described by only two critical exponents. Otherwise, there is
a whole spectrum of critical exponents.\\
\hspace*{0.5cm} The multifractal analysis is more than a simple fit. The $ f-\alpha$ representation is one of the solution of a differential 
equation obtained from a local scale invariance hypothesis [21]. On the other hand, it reflects the details of the avalanche structures. Indeed, the study of various space-time structures of avalanches in one-dimensional local limited sandpile model [22], without bulk dissipation, shows that there are two compact types of space-time avalanches (Fig.1):
 
\begin{figure}
\begin{center}
\includegraphics[width=10cm, height=4cm, angle=0]{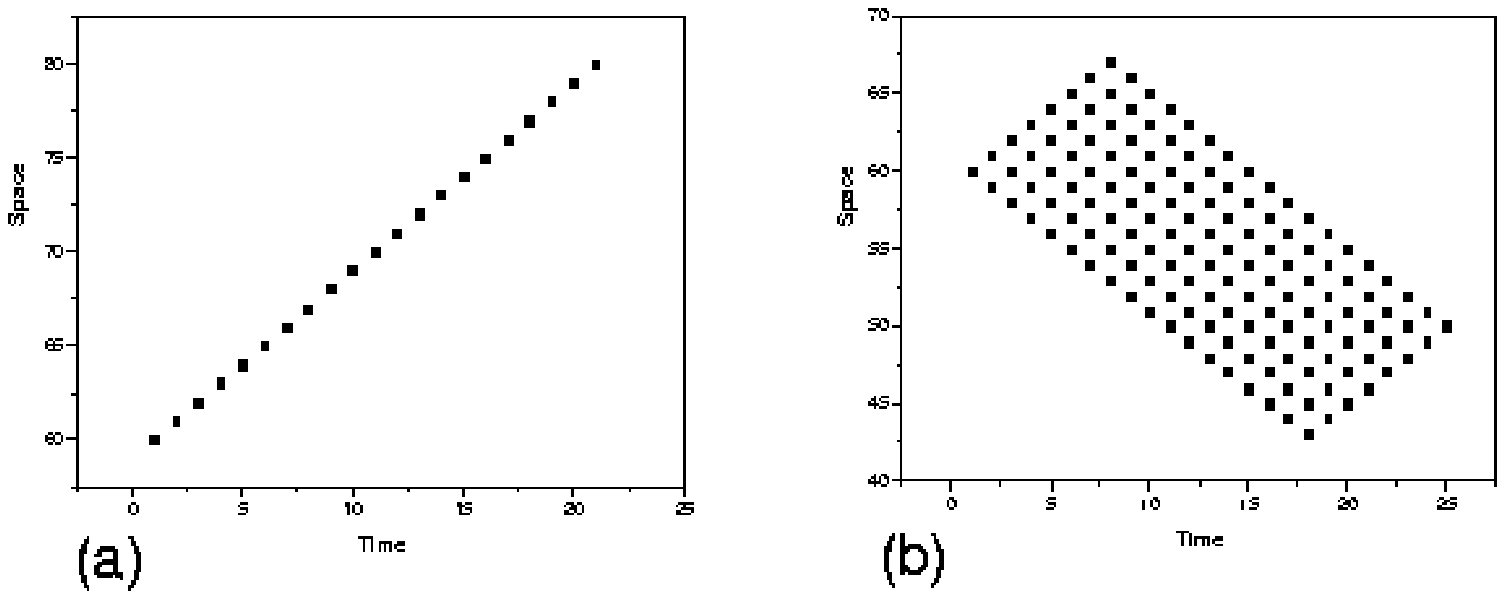}
\caption{The space-time representation of the avalanche for $\epsilon=0$. The dots represent the active sites:(a) example of 1D avalanche, and (b) example of 2D avalanche. Consequently, the system exhibits a multifractal behavior as it presents two different compact structures.}
\end{center}
\end{figure}

1D linear avalanches, and avalanches with backward events which appear to have a 2D space-time structure. The main cause of the lack of self-similarity is the presence of different structures and the effect of finite-size, which under 1D and 2D avalanches is different. It was found that the study of these two types of structures separately gives two different set of critical exponents. As a result, when the scaling is studied with the combined effects of both avalanche types, it gives rise to a spectrum of scaling indices which reveal the presence of multifractal behavior. The multifractal scaling represented by the equation (4) has been found to describe the finite-size effects of the numerically evaluated probability distribution function in a better way.
\section{Dissipation effects on non critical sandpile models}
\hspace*{0.5cm}The models described above have been studied for $\epsilon=0$ [18]. It was shown that they display a multifractal behavior. We
are interested in the effects of the dissipation on these models, and we study whether they can display critical behavior. We conclude by 
using the two techniques described above that for the LL model, the FSS is better for $\epsilon > \epsilon_c=0.04$, while the multifractal analysis,
 represented by Eq. (4), gives a satisfactory fit to our data for smaller values of bulk dissipation. However, for the LU, NLL and NLU models, only
 the multifractal fit is appropriate for all values of  $\epsilon$.

\begin{figure}
\begin{center}
\includegraphics[width=6cm, height=5cm, angle=0]{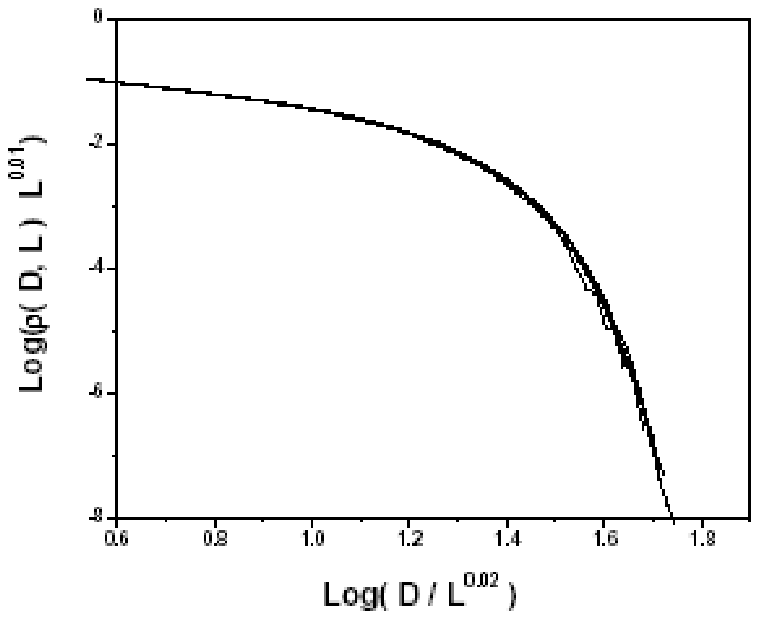}
\caption{The FSS fit of $\rho(D, L)$ for $\epsilon= 0.1$ and  for system 
sizes ranging from L=60 to L=480. We note that the FSS gives a good fit for our data , where $\nu_D=0.02$ and $\beta_D=0.01$.}
\end{center}
\end{figure}

\subsection{Local limited models}
\hspace*{0.5cm}The local limited models were considered as the simplest ones. During each step of an avalanche, the number of grains N which 
falls to the nearest neighbor is kept constant. Since the critical behavior of the system and the avalanche dynamics are independent of the
value of N, we will take, for simplicity, N = 2. In order to have a good fit, we try to adjust the exponents $ \beta$ and $\nu $ for the
FSS analysis, and the parameters $X_0$ and $L_0$ for the multifractal fit, in such way that for each value of dissipation $\epsilon$, the 
distributions $ \rho (F, L) $ and $\rho(D, L)$ overlap for different values of the system size $ L $.\\

\begin{figure}
\begin{center}
\includegraphics[width=6cm, height=9cm, angle=0]{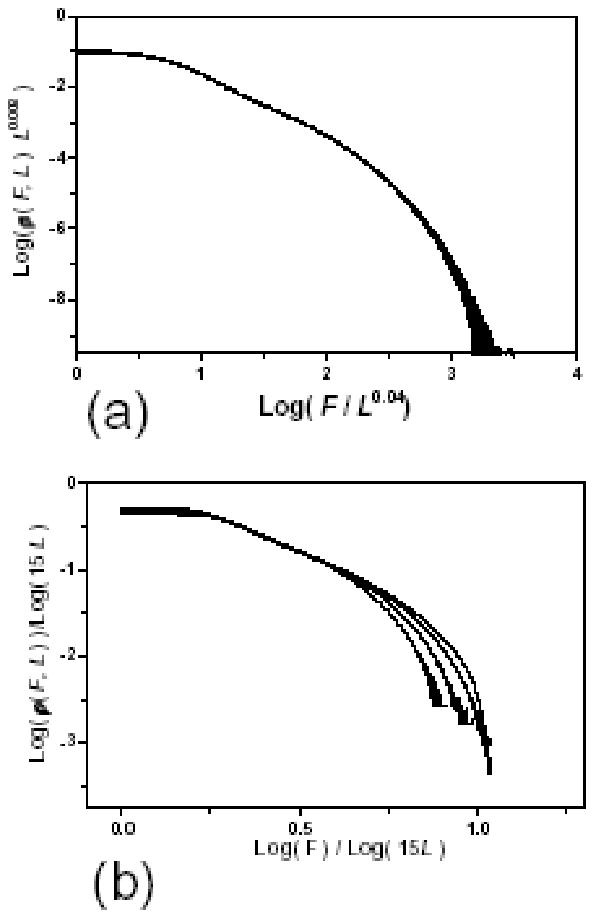}
\caption{The FSS fit of $\rho(D, L)$ for $\epsilon= 0.1$ and  for system sizes ranging from L=60 to L=480. We note that the FSS gives a good fit for our data , where $\nu_D=0.02$ and $\beta_D=0.01$.}
\end{center}
\end{figure}

In fig.2 we present the FSS of the drop number distribution $\rho(D,L)$ for a dissipation rate $\epsilon=0.1$. However, all curves corresponding
to different values of $ L$ overlap within the scaling fit (eq.3), with $\beta_D = 0.01$ and $\nu_D=0.02$. In order to emphasize the fact that
 the FSS gives the best description of the finite size effects, we analyse, in fig.3a, the flip number distribution $\rho(F,L)$ for different
values of the system size $L$. We obtain a good fit for the critical exponents $ \beta _F = 0.008 $ and $\nu_F=0.04$. On the other hand, the 
multifractal analysis of $\rho(F,L)$ (Fig.3b) shows that the data collapses only for small values of $F$ with the exponents $F_0=1$ and
$L_0=1/15$, while for $F_0=25$ and $L_0=1/55$ they overlap only for large values. Thus, we may conclude that for $\epsilon=0.1$ the FSS is
more adequate than the multifractal analysis.\\

\begin{figure}
\begin{center}
\includegraphics[width=10cm, height=8cm, angle=0]{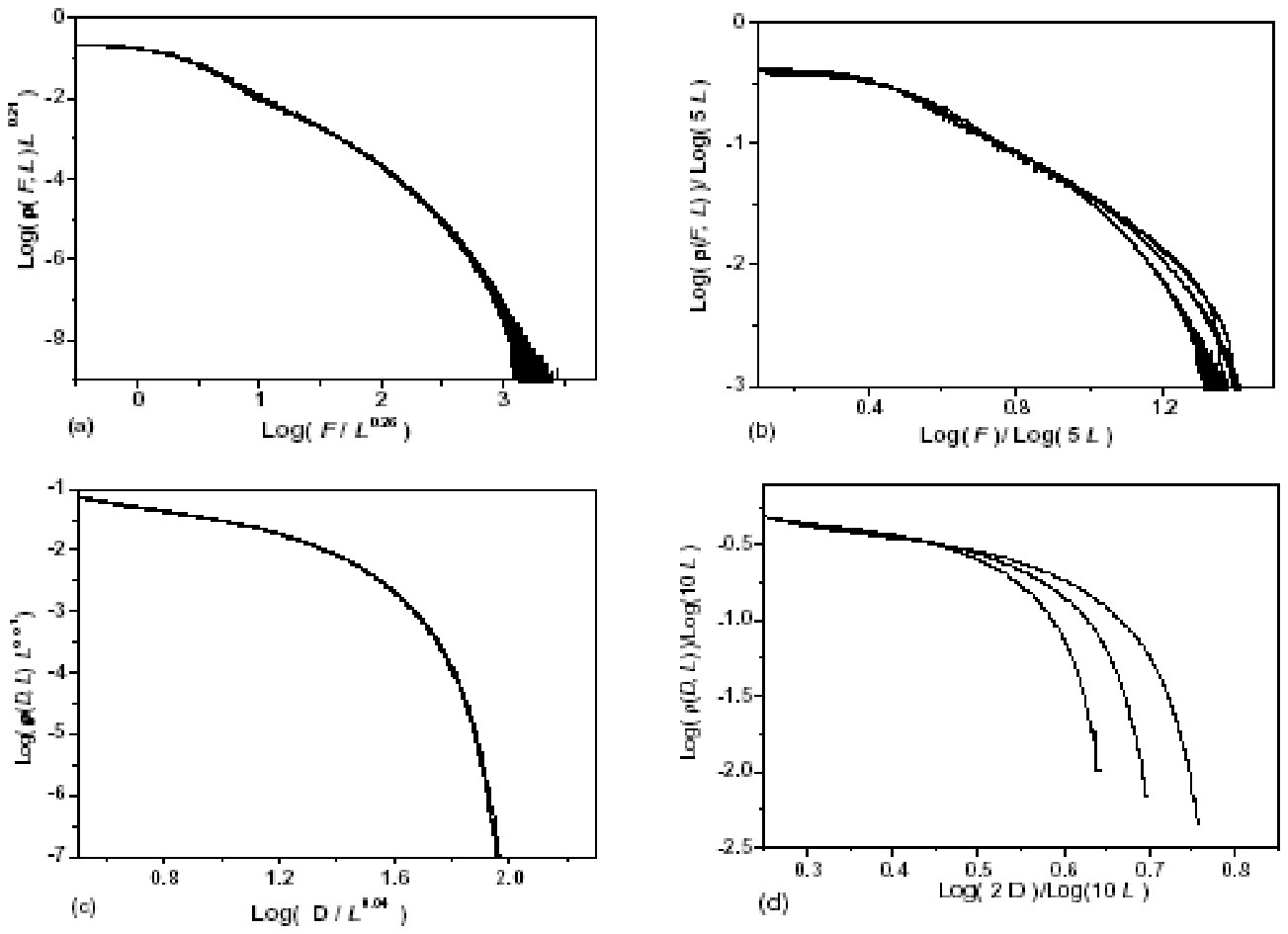}
\caption{(a)-(c) The FSS and (b)-(d) the multifractal analysis of  $\rho(F, L)$ and $\rho(D,L)$, respectively, for  $\epsilon=0.06$ and for system sizes ranging from $L=128$ to $L=512$. The FSS gives the best fit of our data for all values of $\epsilon$ greater than a critical value $\epsilon_c$.}
\end{center}
\end{figure}

\hspace*{0.5cm}In figure 4 the data for both $\rho(F, L)$ and  $\rho(D, L)$ have been plotted for $\epsilon= 0.06$. We find again, as for the previous case which corresponds to $\epsilon= 0.1$, that the FSS form gives an excellent fit over the whole range of the data with $\beta_F=0.21$ and $\nu_F=0.26$ for the flip number (Fig.4a), and $\beta_D=0.01$ and $\nu_D=0.04$  for the drop number (Fig.4c), and the mutlifractal analysis does not give an appreciably better fit. It gives a good fit only for intermediate values of the flip number (Fig.4b) for small values of the drop number (Fig.4d) otherwise the data does not collapse for different sizes $L$. For $\epsilon=0.01$, fig.5a shows that the FSS analysis works well for small values of $F$ but significant discrepancies arise at larger values. There are two different scaling fits which look good over limited range of data. The first one is obtained only for small values of $F$ with $\beta_F=\nu_F=0.38$, while the second one is localized for larger values with $\beta_F=1.7$ and $\nu_F=1.2$. In fig.5b we have plotted the function ${\rm log}_{10}\rho(F,L)/{\rm log}_{10}(18L)$ vs $ {\rm log}_{10}(F)/{\rm log}_{10}(18L)$ for different system sizes $L$. A good fit is obtained since all the data collapse for different values of $L$. In order to emphasize the fact that the multifractal analysis presents the appropriate fit, we show respectively in fig.5c and fig.5d the results of the FSS and the mutifractal analyses for the drop number distribution $\rho(D, L)$. As shown in this figure, the multifractal analysis is a much better way to describe the finite size effects than the analysis based on a simple FSS.\\

\begin{figure}
\begin{center}
\includegraphics[width=9cm, height=7cm, angle=0]{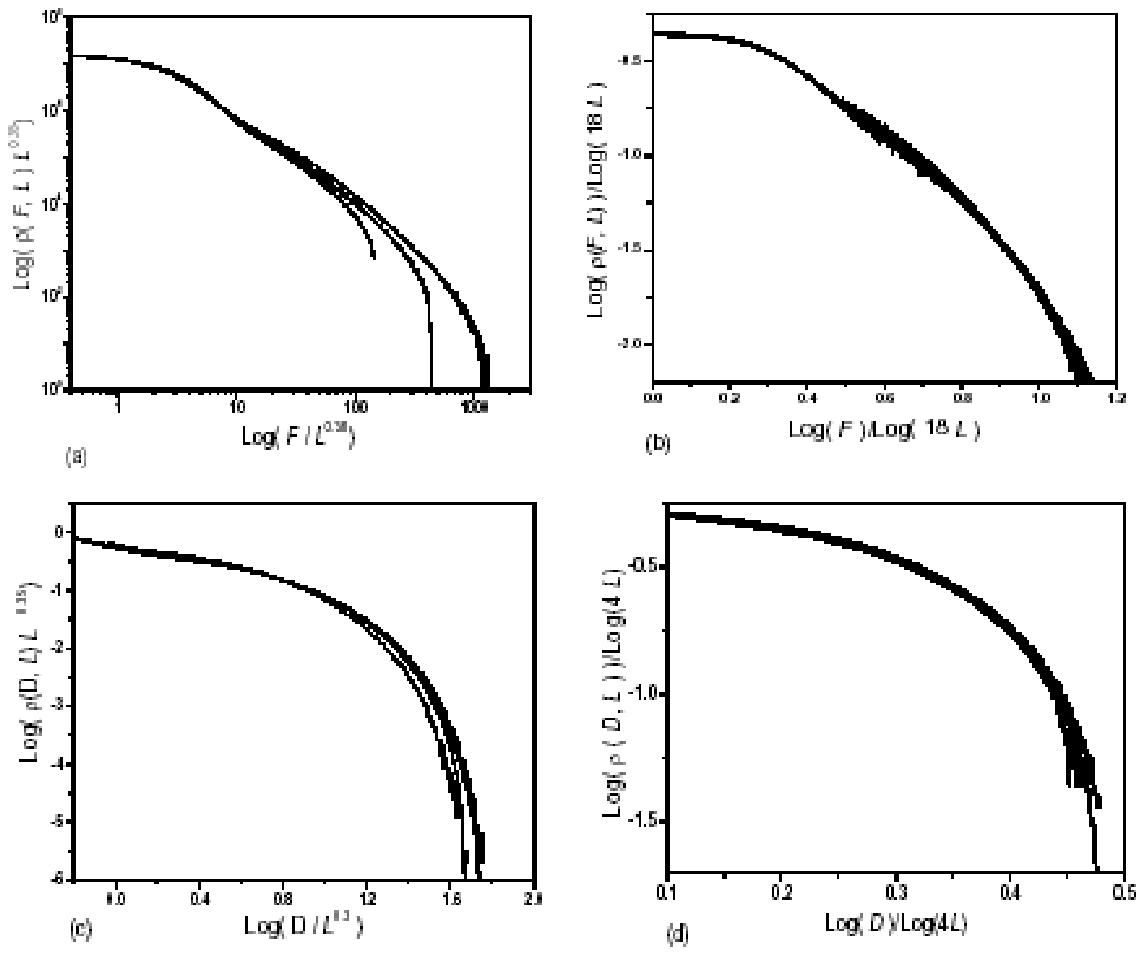}
\caption{(a)-(c) The FSS and (b)-(d) the multifractal analysis of  $\rho(F, L)$ and  $\rho(D,L)$ respectively for  $\epsilon=0.01$, and for system sizes ranging from $L=128$ to $L=512$. The best fit is given by the multifractal analysis due to the fact that for $\epsilon<\epsilon_c$, the structure of avalanches became compact and therefore the 1D and 2D structures begin to react differently.}
\end{center}
\end{figure}

\hspace*{0.5cm}The study of different space-time structures of avalanches for the LL model reveals the existence of two classes. For $\epsilon >\epsilon_c$,
the bulk dissipation has a hitchhiker effect on the avalanches. The active sites frequently dissipate their energy rather than relax to their
nearest neighbor. As a result, the avalanche space-time structures are quasi-two-dimensional with very low compactness(fig.6a-b). Thus, the FSS
analysis gives the best fit because of the quasi "linearity" of the avalanche structure. For $\epsilon <\epsilon_c$ the avalanche structures
are more compact than the previous ones(fig.6c-d). The relaxation is more frequent than the dissipation thereby leading to more backward events that
appears to have a two-dimensional structure. Consequently, the avalanche size distribution obeys multifractal analysis. For both cases,
different structures are generated by varying the dissipation rate $\epsilon$. As a matter of fact, their fractal dimension as well as the
corresponding critical exponents $\beta$ and $\nu$ vary continuously.\\

\begin{figure}
\begin{center}
\includegraphics[width=8cm, height=7cm, angle=0]{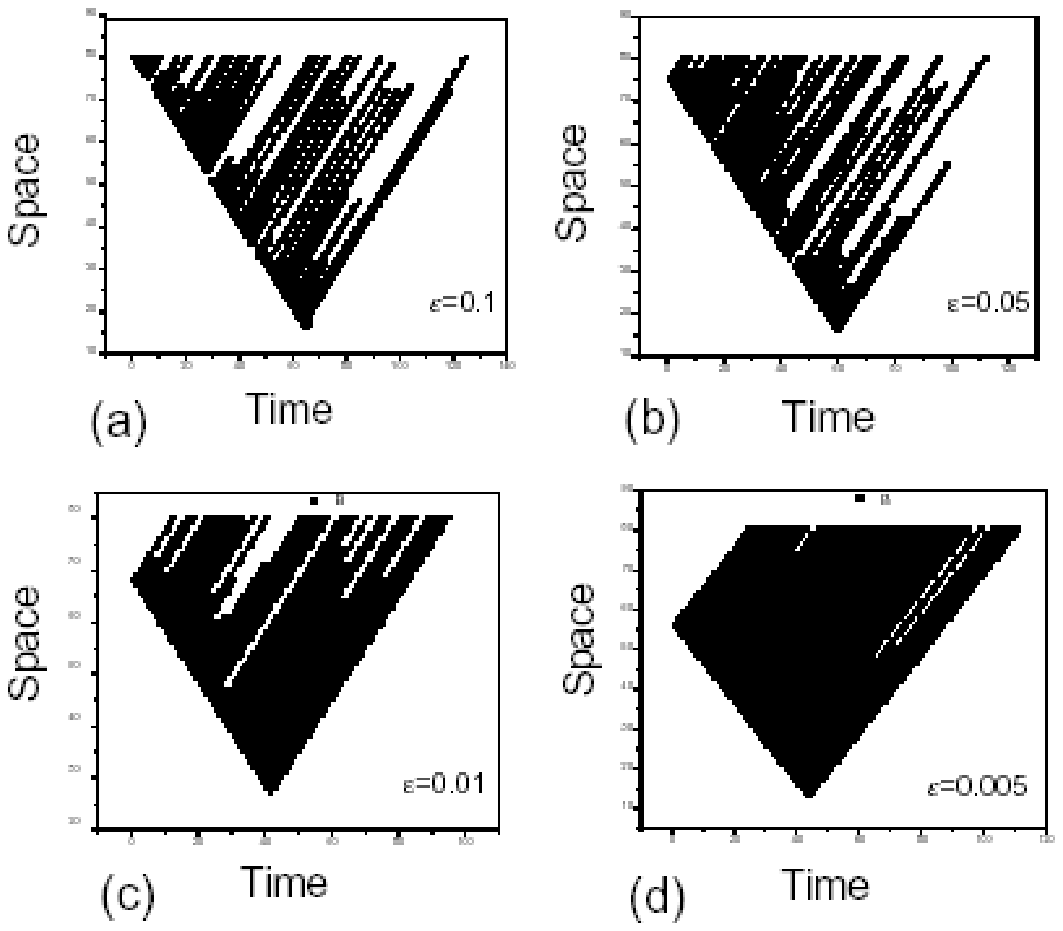}
\caption{By decreasing the level of the dissipation rate from (a) to (d), the structure of avalanches becomes more compact and then the system goes from the situation where the dynamics present a fractal structure, to a multifractal behavior.}
\end{center}
\end{figure}

\begin{figure}
\begin{center}
\includegraphics[width=6.5cm, height=5.5cm, angle=0]{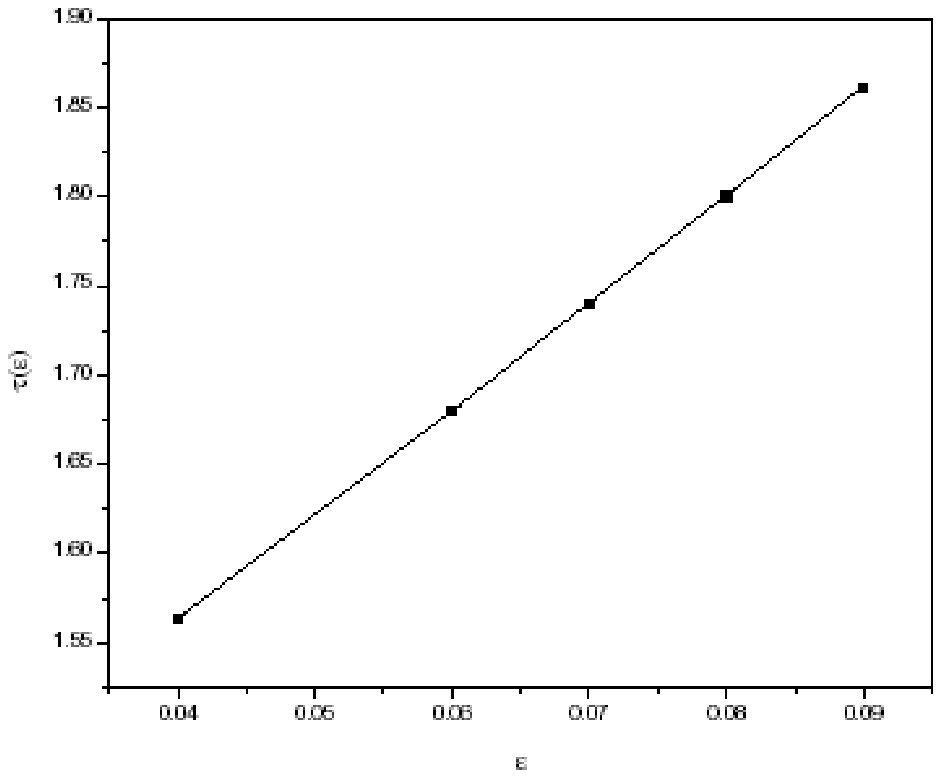}
\caption{The critical exponent $\tau$ versus the dissipation rate for $\epsilon > \epsilon_c$ where $\rho(F)$ presents a power law behavior.}
\end{center}
\end{figure}

\begin{figure}
\begin{center}
\includegraphics[width=6.5cm, height=5.5cm, angle=0]{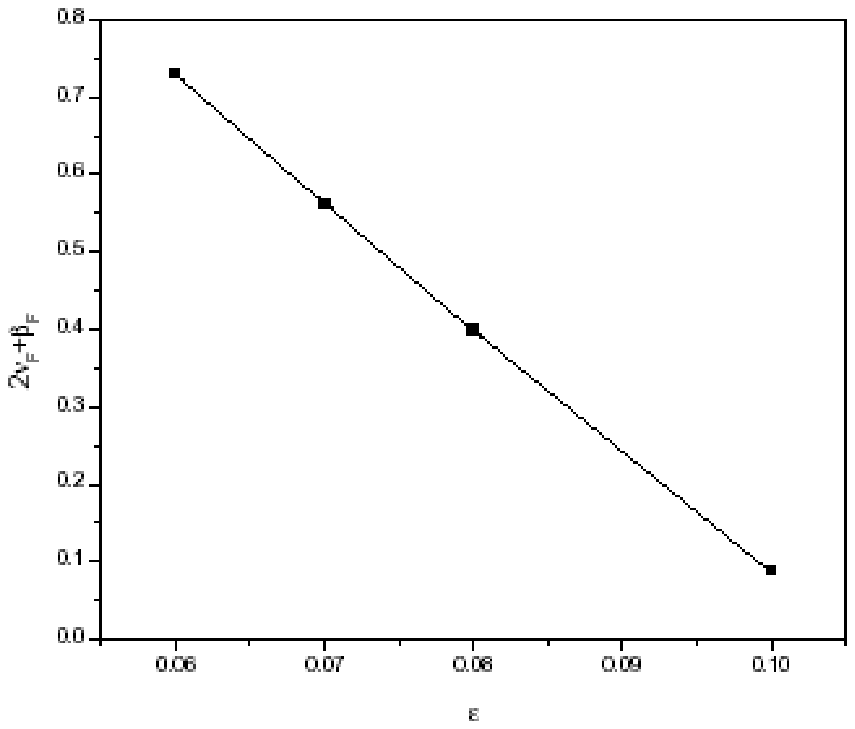}
\caption{The variation of $2\nu_F+\beta_F$ versus the dissipation rate $\epsilon$. The scaling laws have been checked, and show that the system presents a non-universal behavior where the critical exponents are $\epsilon$-dependent.}
\end{center}
\end{figure}

\begin{figure}
\begin{center}
\includegraphics[width=7cm, height=6cm, angle=0]{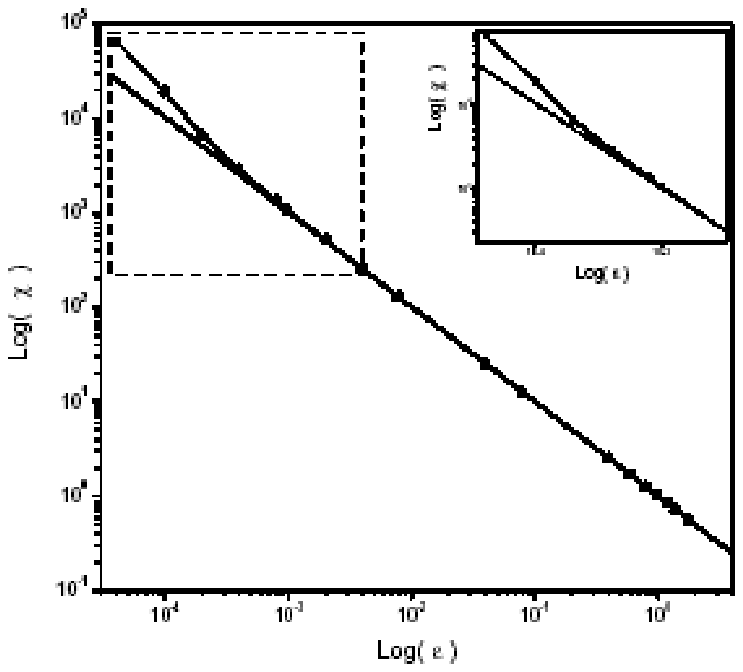}
\caption{The log-log plot of the susceptibility  $\chi(\epsilon) = \partial \rho_a/\partial h_{ext}$ for a system with periodic boundary conditions, and sizes ranging from $L=64$ to $L=256$. The straight line presents the $1/\epsilon$  behavior. It was shown that the susceptibility presents two different variations for low and high values of $\epsilon$. The inset delimits the region where the $1/\epsilon$ behavior does not fit with $\chi$.}
\end{center}
\end{figure}

\begin{figure}
\begin{center}
\includegraphics[width=7cm, height=6cm, angle=0]{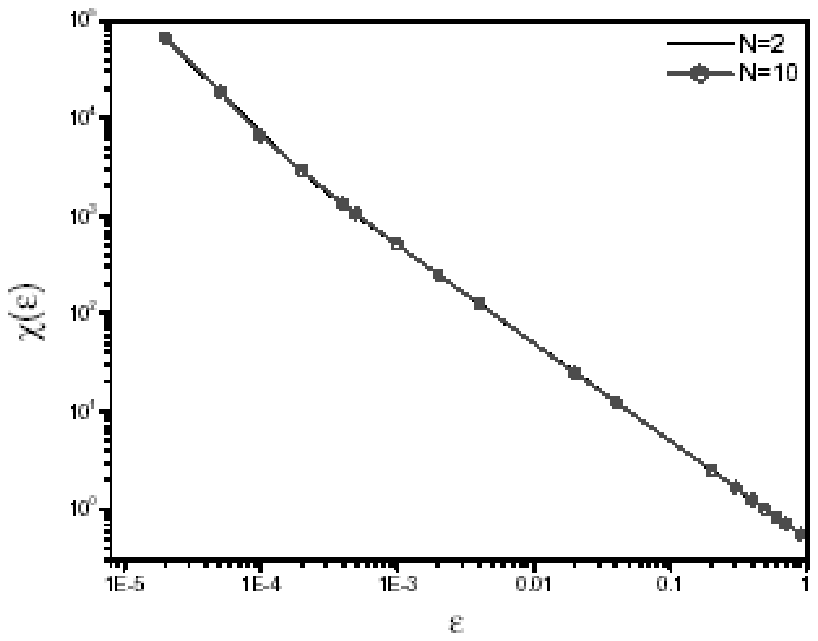}
\caption{The log-log plot of the susceptibility $\chi(\epsilon)$, for different values of the toppling grains N and for a system size $L=128$. As a result, the dynamics of the system is independent of the details of the system.}
\end{center}
\end{figure}

\hspace*{0.5cm} From the FSS and the multifractal techniques, we conclude that the LL model with a bulk dissipation energy exhibits, at some
critical value $\epsilon_c$, a crossover from multifractal to self-similar behavior. In both regions the critical exponents vary continuously
with the dissipation rate $\epsilon$ as the space-time structure of the avalanches changes with $\epsilon$, producing a low compactness
accompanied by a change of the finite-size effects. The distribution function of the flip number presents a power law behavior for high values
of $\epsilon$ and the critical exponent $\tau$ increases linearly with increasing the dissipation rate (Fig.7). The FSS technique was used in
order to check if there are any scaling laws between different critical exponents. As a result, we note that the model presents non-universal 
behavior since the quantity $2\nu_F+\beta_F$ decreases linearly with increasing $\epsilon$(Fig.8). A numerical study of the zero field
 susceptibility $\chi$ ($\chi=\partial \rho_a/\partial h_{ext}$, where $\rho_a$ is the density of active sites) of the LL model with a giving 
value of the toppling grains N, shows that it is singular at vanishing values of $\epsilon$, thereby signalling a long-range response function (Fig.9). 
The critical behavior is recovered in the limit of vanishing driving field, corresponding to the locality breaking in the dynamical evolution.
For large values of the dissipation ( $\epsilon > \epsilon_c$), the zero field susceptibility exhibits a power law behavior, $1/\epsilon$, similar to usual sandpile models. For $\epsilon \rightarrow 0$, it presents a faster increase than a simple power law, thereby signalling a more 
complicated behavior. We mention that for $\epsilon < \epsilon_c$, unlike the usual sandpile models, the density of active sites $\rho_a$ 
does not vary linearly with the external parameter $h_{ext}$. The behavior of $\chi$ vs $\epsilon$ has been studied for different values of the
toppling grains number N (Fig.10), and it's shown that the crossover behavior is insensitive to the details of the system in the case of LL 
models.
\subsection{Local unlimited and non local models}
\hspace*{0.5cm}The investigations of the LU, NLL and NLU models shows that the multifractal analysis is the most appropriate technique, for 
all values of  $\epsilon$, and gives a good fit for different system sizes. On the other hand, the zero field susceptibility $\chi(\epsilon)$ 
does not exhibit $1/\epsilon$-like power law behavior. It displays rather different behavior characterized by some function $f(\epsilon)$ that diverges (at vanishing values of $\epsilon$) more quickly than any power law behavior. The variation of the number of grains N that flip for any relaxation process, and the non locality presented by these models, preserve the non critical power law behavior even for high values of the dissipation rate. Since the dynamics of these models under variable and non local rules generate back avalanches that are responsible for an appropriate multifractal fit, we believe that the dissipation can neither eliminate nor reduce them considerably. Thus, a non power law critical behavior of the susceptibility is proposed. The critical exponents that give the appropriate multifractal fit vary continuously by varying $\epsilon$.

\section{Conclusion}
\hspace*{0.5cm}To summarize, we have studied the behavior of one dimensional sandpile models with bulk dissipation. The various distribution
functions of events, for each value of the dissipation rate $\epsilon$, depend on the system size. However, an investigation of the scaling
properties of the one dimensional LL, LU, NLL and NLU models have been established using both simple FSS and multifractal analyses of our data.
Using both FSS and multifractal techniques, we have shown that for the LL model the first one works better for high values of the dissipation
rate $\epsilon > \epsilon_c$, while the second gives a more appropriate fit for low values of $\epsilon$. The analysis of the space-time structure
of avalanches shows that such structures are quasi-two-dimensional for high values of $\epsilon$, whereas they become more compact 2D areas for
low values. Consequently, the self-similarity breaks down for $\epsilon \leq \epsilon_c$, leading to multifractal behavior resulting
 from two kinds of space-time structures (1D and 2D compact avalanches) which start to react differently. In order to emphasize the result
obtained using the techniques mentioned above, the crossover behavior has been checked by the calculation of the zero field susceptibility
$\chi(\epsilon)$ in view of the fact that it exhibits $1/\epsilon$ -like power law behavior for large values of $\epsilon$, while for small
 values such behavior is replaced by a rather different one. We note that since the crossover behavior is localized at very low values of
$\epsilon$, and that the system deviates slowly from the power law behavior, the susceptibility, the FSS and multifractal techniques
do not allow high precision of the estimated value $\epsilon_c=0.04$.\\
The LU, NLL and NLU models do not display any crossover behavior, and they preserve the multifractal analysis since their intrinsic dynamics produce
frequent back avalanches (for all values of $\epsilon$) that are responsible for such behavior. The critical exponents obtained either within
the multifractal analysis or the FSS vary continuously with the value of the dissipation rate $\epsilon$.

\vspace*{1.5cm}
{\bf Aknowlendgements}\\
We thank D. Dhar for clarifying the significance of SOC and its connection to the multifractal and the self-similar behaviors.
The authors are grateful to the high education ministry MESFCRS for the financial support in the framework of the program PROTARSIII,
grant no: D12/22.
 
\vspace*{3cm}
\section*{References}
\begin {enumerate}
\item[{[1]}]P. Bak, C. Tang, and K. Wiesenfield, Phys. Rev. A {\bf 38}, 364 (1988).
\item[{[2]}]C. Tang and P. Bak, Phys. Rev. Lett. {\bf 60}, 2347 (1988); J. Stat. Phys. {\bf 51}, 797 (1988).
\item[{[3]}]P. Bak and K. Chen, Physica D {\bf 38}, 5 (1989).
\item[{[4]}]K. Wiesenfield, C. Tang, and P. Bak, J. Stat. Phys. {\bf 54}, 1441 (1989).
\item[{[5]}]Bak, P., {\it How nature works: the science of self-organized criticality} (Springer Verlag,1996).\\
H. J. Jensen,{\it SelfOrganized Criticality: Emergent Complex Behavior in Physical and Biological Systems}, Cambridge university press (1998).
\item[{[6]}]A. Vespignani and S. Zapperi, Phys. Rev. E {\bf 57}, 6345 (1998).
\item[{[7]}]R. Pastor-satorras and A. Vespignani, Phys. Rev. E {\bf 62}, 6195 (2000).
\item[{[8]}]H.J. Ruskin, Y. Feng, Physica A {\bf 245}, 453 (1997).
\item[{[9]}]A. Benyoussef, A. El Kenz, M. Khfifi, and M. Loulidi, Phys. Rev. E {\bf 66}, 041302(2002).\\
S. L\"ubeck, and K. D. Usadel, Fractals {\bf 1}, 1030(1993).\\
S. L\"ubeck, and D. Dhar, Journal of Statistical Physics {\bf 102}, 1 (2001).
\item[{[10]}]A. Vespignani and S. Zapperi, Phys. Rev. Lett. {\bf 78}, 4793 (1997).
\item[{[11]}]G. Grinstein in {\it "Scale Invariance, Interfaces and Nonequilibrium Dynamics"}, NATO Advanced Study Institute, Series B: Physics, volume {\bf 344}, A. McKane et al. Eds. (Plenum,NY, 1995).
\item[{[12]}]D. Dhar and R. Ramaswamy, Phys. Rev. Lett. {\bf 63}, 1659 (1989).\\
D. Dhar , Phys. Rev. Lett. {\bf 64}, 1613 (1990).Physica A  {\bf 263}, 4 (1999).\\
S. S. Manna, J. Phys. A{\bf 24}, L363 (1991).\\
E.V. Ivashkevich, V.B. Priezzhev, Physica A  {\bf 254}, 97 (1998).
\item[{[13]}]H. M. Jaeger, Chu-heng Liu, and S. R. Nagel, Phys. Rev. Lett. {\bf 62}, 40 (1989).
\item[{[14]}]V. Frette, K. Christensen, A. Malthe-Sorensen, J. Feder, T. Jossang, and P. Meakin, Nature (London) {\bf 376}, 49 (1996).
\item[{[15]}]D. Dhar , e-print Cond-mat/9909009 and references therein.
\item[{[16]}]R. Dickman, M. A. Muñoz, A. Vespignani, and S. Zapperi, Braz. J. Phys. {\bf 30}, 27 (2000).
\item[{[17]}]R. Dickman, A. Vespignani, and S. Zapperi, Phys. Rev. E {\bf 57}, 5095 (1998).\\
A. Vespignani, R. Dickman, M.A. Munoz, and Stefano Zapperi, Phys. Rev. Lett. {\bf 81}, 5676 (1998).
\item[{[18]}]L.P. Kadanoff, S.R. Nagel, L. Wu, and S.M. Zhou, Phys. Rev. A {\bf 39}, 6524 (1989).
\item[{[19]}]C. Tebaldi, M. De Menech, and A. L. Stella , Phys. Rev. Lett. {\bf 83}, 3952 (1999).
\item[{[20]}]T. C. Halsey, M. H. Jensen, L. P. Kadanoff, I. Procaccia, and B. I. Shraimen, Phys. Rev. A {\bf 33}, 1141(1986).\\
T. C. Halsey, P. Meakin, and I. Procaccia, Phys. Rev. Lett {\bf 56}, 854(1986).
\item[{[21]}]C. Tang, "Scalings in Avalanches and Elsewhere", ITP Preprint (NSF-ITP-89-118), unpublished.
\item[{[22]}]B. A. Carreras, V. E. Lynch, D. E. Newman, and R. Sanchez, Phys. Rev. E {\bf 66}, 011302 (2002).
\end{enumerate}

\end{document}